\documentclass[preprintnumbers,aps,prd,longbibliography,nofootinbib,nobibnotes,showpacs,amsmath,amssymb]{revtex4-1}
\usepackage{amsfonts}
\usepackage{mathrsfs}
\usepackage{epsfig}
\usepackage{graphicx}%
\usepackage{dcolumn}
\usepackage{amsmath}
\usepackage{color}
\usepackage{subfigure}

\begin{document}
\renewcommand{\baselinestretch}{1.3}
\newcommand\beq{\begin{equation}}
\newcommand\eeq{\end{equation}}
\newcommand\beqn{\begin{eqnarray}}
\newcommand\eeqn{\end{eqnarray}}
\newcommand\nn{\nonumber}
\newcommand\fc{\frac}
\newcommand\lt{\left}
\newcommand\rt{\right}
\newcommand\pt{\partial}

\title{Extending Bekenstein's theorem in order to search  exact solutions of Einstein-Maxwell-conformal-scalar equations }

 \author{Jianhui Qiu}\email{jhqiu@nao.cas.cn}
\author{Changjun Gao}\email{gaocj@nao.cas.cn}
\affiliation{ National Astronomical Observatories, Chinese Academy of Sciences, Beijing {100101}, China}

\affiliation{School of Astronomy and Space Sciences, University of Chinese Academy of Sciences,
No. 19A, Yuquan Road, Beijing 100049, China}

\begin{abstract}
 The Bekenstein's theorem allows us to generate a Einstein-conformal scalar solution from a single Einstein-ordinary scalar solution. In this article, we extend this theorem to  Einstein-Maxwell-scalar (EMS) theory with a non-minimal coupling between the scalar and Maxwell field. As  applications of this extended theorem, the well-known static dilaton solution and rotating solution with a specific coupling between dilaton and Maxwell field are considered, and new conformal dilaton black hole solutions are found. The Noether charges such as the mass, electric charge, angular momentum are compared between the old and new black hole solutions connected by conformal transformations, and they are found conformally invariant. We speculate that the theorem may be helpful in the computations of metric perturbations and spontaneous scalarization of black holes in the Einstein-Maxwell-conformal-
scalar theory since they can be mapped to the corresponding EMS theories, which have been investigated in  detail.

\end{abstract}

\keywords{1;2}

% \Keywords{ }

%11.10.Kk Field theories in dimensions other than four (see also 04.50.-h Higher-dimensional gravity and other theories of gravity; 04.60.Kz Lower dimensional models; minisuperspace models in general relativity and gravitation)

%04.50.Kd 	Modified theories of gravity

%04.50.-h Higher-dimensional gravity and other theories of gravity
%         (see also 11.25.Mj Compactification and four-dimensional models, 11.25.Uv D branes)

% 04.50.+h Gravity in more than four dimensions, Kaluza-Klein theory,
% unified field theories, alternative theories of gravity
%(see also 11.25.M Compactification and four-dimensional models), dilaton gravity

% 04.50.Gh Higher-dimensional black holes, black strings, and related objects.

% 04.70.Dy Quantum aspects of black holes, evaporation, thermodynamics.

% 11.27.+d Extended classical solutions; cosmic strings,
%domain walls, texture (see 98.80.C in cosmology)
%98.80.-k Cosmology
%\keywords{}
\pacs{04.50.Kd, 04.70.Dy}
\maketitle


%%%%%%%%%%%%%%%%%%%%%%%%%%%%%%%%%%%%%%%%%%%%%%%%%%%%%%%%%%%%%%%%%%%%%%%%%%%%%%%%%%%%%%%

\section{Introduction}
As is known to all, it is very difficult to find exact solutions to Einstein's equations in General Relativity, let alone the exact solutions to Einstein's equations with a conformal scalar stress-energy. In 1974, Bekenstein \cite{bekenstein1974exact} presented a theorem by means of which one can generate Einstein-Maxwell-conformal-scalar (EMCS) solutions from a  solution of the coupled Einstein-Maxwell-ordinary scalar theory, wherein the scalar field is minimally coupled to the Maxwell invariant. But he didn't take into account the non-minimal coupling between the matter fields, e.g.  the coupling between the scalar field and the Maxwell field in Einstein's frame. We know this kind of coupling is very natural theoretically, as shown by string theory. Then if one goes further, namely considering this situation, can Bekenstein's theorem be generalized from the EMS theory with minimal couplings to EMS with non-minimal couplings between the scalar and Maxwell fields?  The answer is ``yes''. In this article, we shall conduct this generalization. \footnote{Unless otherwise stated, ``EMS'' stands for Einstein-Maxwell-scalar theory  with a non-minimal coupling between the scalar and Maxwell fields.}

EMS theory has recently triggered much interest due to the discovery of the spontaneous scalarisation phenomenon, which was mainly studied in the scalar-tensor-Gauss-Bonnet gravity \cite{doneva2018new,silva2018spontaneous,antoniou2018evasion}. This fascinating phenomena is the only known dynamical method for endowing black holes (and other compact objects) with scalar hair without affecting the weak field predictions. It is found that, in gravitational models where a real scalar field non-minimally couples to the  Gauss-Bonnet invariant $f(\varphi) \mathscr{G}$
 and under certain choices of the coupling function, both the standard (bald) vacuum solutions of General Relativity and new ``hairy" BH solutions with a scalar field profile are possible. Thus, it circumvents the black hole no-hair theorems(see\cite{sotiriou2015black,herdeiro2015asymptotically}for reviews). For some range of mass, a Schwarzschild BH becomes unstable and transfers some of its energy to a ``cloud'' of scalar particles around it. This spontaneous scalarisation is triggered by the strong spacetime curvature, which induces non-linear curvature terms in the evolution equations. As a result, the computations are intensive and complicate \cite{minamitsuji2019scalarized,silva2019stability}.

However, the spontaneous scalarisation with dynamical process could be confirmed in a cousin model: EMS theory \cite{herdeiro2018spontaneous}. In this class of models, there is no coupling between the scalar field and the curvature, but a non-minimal coupling between the
scalar and the electromagnetic field is present. 
For certain values of the coupling function , the conventional electrovacuum (scalar-free) Reissner– Nordström (RN) BH solves the equations of motion in EMS models, as does a novel class of BHs that allows for a non-trivial scalarfield configuration (scalar-hair).
When the charge-mass ratio of the RN BH is sufficiently high, the RN BH becomes unstable to scalar perturbations, and the formation of these hairy BHs is hypothesized to represent the endpoint of the instability\cite{fernandes2019charged,fernandes2019spontaneous}. This technically simpler model facilitated the investigation of the domain of solutions' existence, particularly outside the spherical sector, as well as the performance of completely non-linear dynamical evolutions.
For more information on the spontaneous scalarization of charged black holes in EMS, one can refer to \cite{herdeiro2018spontaneous,fernandes2019spontaneous,astefanesei2019einstein}. In view of these points, it is necessary to extend  Bekenstein's theorem from the EMS theory with  minimal couplings to EMS with non-minimal couplings .

The paper is organized as follows. In Section II, we show the actions for a generic EMS theory and the conformal version first, and then explore how they can be connected by the theorem and the strict proof is provided. In Section III, we focus on the Einstein-Maxwell-dilaton theory \cite{horne1992rotating} as an example to interpret the theorem. The conformal solution and corresponding thermodynamics are derived. In Section IV, we concentrate on the rotating dilaton solution in the case of $K_1{(\phi)}=e^{-2\sqrt{3}\phi}$, which is first achieved by the  Kaluza-Klein black hole solution \cite{frolov1987charged}. Thanks to this extended theorem, the conformal rotating dilaton solution can be achieved straightforwardly. We investigate the first law of thermodynamics and find that all the quantities are unchanged under conformal transformation. In section V, inspired by the dilaton theory, we prove that for general EMS theory, all the physical  quantities in the first law of thermodynamics are all conformally invariant provided that the black holes are asymptotically flat and the scalar field asymptotically vanishes at infinity. In section VI, we give the conclusion and discussion.

\section{EXTENDING BEKENSTEIN'S THEOREM}
Maxwell's equations are known to be conformally invariant in Einstein's frame, while ordinary massless scalar equations are not. However, there does exist a conformally invariant scalar equation:
 \begin{equation}
 \nabla^{\mu} \nabla_{\mu} \psi-\frac{R \psi}{6}=0\;.
 \end{equation}
 To get exact solutions of Einstein's equations with a conformal scalar stress-energy, Bekenstein \cite{bekenstein1974exact} presented a theorem by means of which one can generate Einstein-conformal scalar solutions from a single Einstein-ordinary scalar solution. Nevertheless, he did not consider the non-minimal coupling between the scalar field and electromagnetic field. The coupling can naturally emerge in many physics, for example in string theory. Now we extend his theorem by considering this coupling.

The Einstein equations for EMS  can be derived from the action principle \cite{herdeiro2020electromagnetic}
 \begin{equation}
 \mathcal{S}_{1}=\frac{1}{16 \pi} \int d^{4} x  \sqrt{-\bar{g}}\left[\bar{R}-2 \bar{\nabla}_{\mu} \phi \bar{\nabla}^{\mu} \phi-K_{1}(\phi) \bar{F}_{\mu\nu}\bar{F}^{\mu\nu}-L_{1}(\phi) \bar{F}_{\mu\nu}
   {} ^{ *} \bar {F} ^{\mu\nu} \right]\;.
 \end{equation}
 Here $\bar{g}$, $\bar{\nabla}$ and $\bar{R}$ represent the determinant, compatible derivative operator and scalar curvature of the metric $\bar{g}_{\mu\nu}$,  respectively. Maxwell field and its Hodge dual are denoted by $\bar{F}_{\mu\nu}$ and $^{*}\bar{F}_{\mu\nu}$. $K_{1}(\phi)$  and $L_{1}(\phi)$ represent the coupling between scalar field and Maxwell field. We note that the quantities in action $\mathcal{S}_1$ are designated with the symbol of bar, and the corresponding quantities labeled with tilde in the following $\mathcal{S}_{2}$  share the same meanings as those in $\mathcal{S}_{1}$. The quantities without labeling a bar or a tilde have a universal meaning. For instance, $ G_{\mu \nu}\left(\bar{g}_{\mu\nu}\right)$ denotes the Einstein tensor with respect to $\bar{g}_{\mu\nu}$, whereas  $G_{\mu \nu}\left(\tilde{g}_{\mu\nu}\right)$ denotes the Einstein tensor  with respect to $\tilde{g}_{\mu\nu}$.

Variation of $ \mathcal{S}_{1}$ with respect to $\bar{g}_{\mu\nu}$ gives

 \begin{equation}
 G_{\mu \nu}\left(\bar{g}_{\mu\nu}\right)=2\left\{S_{\mu \nu}\left[\bar{g}_{\mu\nu},\phi\right]+\tau_{\mu\nu}\left[\bar{g}_{\mu\nu},K_1(\phi),L_{1}(\phi),\bar{F}_{\mu\nu}\right]\right\}\;,
 \end{equation}

 where
 \begin{equation}
 \label{emt1}
S_{\mu \nu}\left[\bar{g}_{\mu\nu},\phi\right]=-\frac{1}{2}\bar{\nabla}_{\gamma} \phi \bar{\nabla}^{\gamma} \phi \bar{g}_{\mu \nu}+ \bar{\nabla}_{u} \phi \bar{\nabla}_{\nu} \phi\;,
 \end{equation}

 and
 \begin{equation}
 \label{tau1}
\tau_{\mu\nu}\left[\bar{g}_{\mu\nu},K_{1}(\phi),L_{1}(\phi),\bar{F}_{\mu\nu}\right]=K_{1}(\phi)\left(\bar{F}_{\mu\gamma}\bar{F}_{\nu}{}^{\gamma}-\frac{\bar{F}^2}{4}\bar{g}_{\mu\nu}\right)+L_{1}(\phi)\left(\bar{F}_{\mu\gamma} {}^{*} \bar {{F}}_{\nu}{}^{\gamma}-\frac{\bar{F}_{\alpha \beta} {}^{*} \bar {F} ^{\alpha \beta}}{4}\bar{g}_{\mu\nu}\right)\;.
 \end{equation}
On the other hand, the Einstein equations for EMCS can be derived from the action
 \begin{equation}
 \mathcal{S}_{2}=\frac{1}{16 \pi} \int d^{4} x  \sqrt{-\tilde{g}}\left[\tilde{R}-2 \tilde{\nabla}_{\mu} \psi \tilde{\nabla}^{\mu} \psi-\frac{\tilde{R}}{3}\psi^2-K_{2}(\psi)\tilde {F}_{\mu\nu} \tilde{F}^{\mu\nu}-L_{2}(\psi) \tilde{F}_{\mu\nu} {}^{*} {\tilde{F}} ^{\mu\nu}\right]\;,
 \end{equation}
 and it reads
  \begin{equation}
  \label{ein2}
 G_{\mu \nu}\left(\tilde{g}_{\mu\nu}\right)=2\left[\Theta	_{\mu \nu}\left[\tilde{g}_{\mu\nu},\psi\right]+\tau_{\mu\nu}\left[\tilde{g}_{\mu\nu},K_2(\psi),L_2(\psi),\tilde{F}_{\mu\nu}\right]\right]\left(1-\frac{1}{3}\psi^2\right)^{-1}\;,
 \end{equation}

where

\begin{equation}
\Theta	_{\mu \nu}\left[\tilde{g}_{\mu\nu},\psi\right]=S_{\mu \nu}\left[\tilde{g}_{\mu\nu},\psi\right]-\frac{1}{6}\tilde{\nabla}_{\mu}\tilde{\nabla}_{\nu}\psi^2+\tilde{g}_{\mu\nu}\tilde{\nabla}_{\alpha}\tilde{\nabla}^{\alpha}\psi^2\;,
\end{equation}

\begin{equation}
S_{\mu \nu}\left[\tilde{g}_{\mu\nu},\psi\right]=-\frac{1}{2}\tilde{\nabla}_{\gamma} \psi \tilde{\nabla}^{\gamma} \psi \tilde{g}_{\mu \nu}+ \tilde{\nabla}_{\mu} \psi \tilde{\nabla}_{\nu} \psi\;,
\end{equation}

\begin{equation}
\label{tau2}
\tau_{\mu\nu}\left[\tilde{g}_{\mu\nu},K_2(\psi),L_2(\psi),\tilde{F}_{\mu\nu}\right]=K_{2}(\psi)\left(\tilde{F}_{\mu\gamma}\tilde{F}_{\mu\nu}{}^{\gamma}-\frac{\tilde{F}^2}{4}\tilde{g}_{\mu\nu}\right)+L_{2}(\psi)\left(\tilde{F}_{\mu\gamma} {}^{*} {\tilde{F}}_{\nu}{}^{\gamma}-\frac{\tilde{F}_{\alpha \beta} {}^{*} {\tilde{F}} ^{\alpha \beta}}{4}\tilde{g}_{\mu\nu}\right)\;.
\end{equation}

Theorem: If $\bar{g}_{\mu\nu}$, $\phi$ , $\bar{F}_{\mu\nu}$ form a solution of Einstein's equations described by action $\mathcal{S}_{1}$, then $\tilde{g}_{\mu\nu}=\Omega^{-2}\bar{g}_{\mu\nu}$, $\psi=\sqrt{3}\tanh \frac{\phi}{\sqrt{3}}$ and $\tilde{F}_{\mu\nu}=\bar{F}_{\mu\nu}$ with $\Omega^{-1}=\cosh \frac{\phi}{\sqrt{3}}$ form a solution for the EMCS theory described by action $\mathcal{S}_{2}$ provided that $K_{2}(\psi)=K_{2}\left(\sqrt{3}\tanh \frac{\phi}{\sqrt{3}}\right)\equiv K_{1}(\phi)$.

\emph{Proof}: Under the conformal transformation
\begin{equation}
\bar{g}_{\mu\nu}=\Omega^2\tilde{g}_{\mu\nu}\;,
\end{equation}
the Einstein tensor transforms as
\begin{equation}
\label{ein3}
G_{\mu\nu}\left[\tilde{g}_{\mu\nu}\right]=G_{\mu\nu}\left[\bar{g}_{\mu\nu}\right]-6\Omega^{-2}\Theta	_{\mu \nu}\left[{\tilde{g}}_{\mu\nu},\Omega\right]\;.
\end{equation}
It is obvious that we have from (\ref{tau1})
\begin{equation}
\label{tau3}
\tau_{\mu\nu}\left[\bar{g}_{\mu\nu},K_{1}(\phi),L_{1}(\phi),\bar{F}_{\mu\nu}\right]=\Omega^{-2}\tau_{\mu\nu}\left[\tilde{g}_{\mu\nu},K_1(\phi),L_{1}(\phi),\tilde{F}_{\mu\nu}
\right]\;.
\end{equation}
It follows then from (\ref{emt1})
\begin{equation}
\label{emt2}
S_{\mu \nu}\left[\bar{g}_{\mu\nu},\phi\right]=S_{\mu \nu}\left[\tilde{g}_{\mu\nu},\phi\right]=\left(\frac{\partial \phi}{\partial \psi}\right)^2S_{\mu \nu}\left[\tilde{g}_{\mu\nu},\psi\right]\;.
\end{equation}
If we let
\begin{equation}
\Omega^2=1-\frac{1}{3}\psi^2\;,
\end{equation}
and regard $\psi$ as some unspecified function of $\phi$, then equation (\ref{ein3}) reduces to
\begin{equation}
\begin{aligned}
\label{ein4}
G_{\mu\nu}\left[\tilde{g}_{\mu \nu}\right]=& 2\left\{\left[\left(\frac{\partial \phi}{\partial \psi}\right)^{2}-\left(1-\frac{\psi^{2}}{3}\right)^{-2} \frac{\psi^{2}}{3}\right] S_{\mu \nu}\left[\tilde{g}_{\mu\nu}, \psi\right]\right.\\
&\left.+\Omega^{-2}\left[-\frac{1}{6} \tilde{\nabla}_{u} \tilde{\nabla}_{\nu} \psi^{2}+\frac{\tilde{g}_{\mu\nu}}{6} \tilde{\nabla}^{\gamma} \tilde{\nabla}_{\gamma} \psi^{2}+\tau_{\mu\nu}\left[\tilde{g}_{\mu\nu}, K_{1}(\phi),L_{1}(\phi), \tilde{F}_{u \nu}\right]\right]\right\}\;,
\end{aligned}
\end{equation}
in view of equation (\ref{tau3}) and (\ref{emt2}).

Assuming $\psi$ satisfies
\begin{equation}
\label{eqpsi}
\left(\frac{\partial \psi}{\partial \phi}\right)^{2}=\left(1-\frac{\psi^{2}}{3}\right)^{2}\;,
\end{equation}
which can be solved by
\begin{equation}
\psi=\sqrt{3}\tanh \frac{\phi}{\sqrt{3}}\;,
\end{equation}
and
\begin{equation}
K_{2}(\psi)=K_{2}\left(\sqrt{3}\tanh \frac{\phi}{\sqrt{3}}\right)\equiv K_{1}(\phi)\;,\ \ \ L_{2}(\psi)=L_{2}\left(\sqrt{3}\tanh \frac{\phi}{\sqrt{3}}\right)\equiv L_{1}(\phi)\;,
\end{equation}
\begin{equation}
\Omega^2=1-\frac{1}{3}\psi^2=1-\tanh^2 \frac{\phi}{\sqrt{3}}=\frac{1}{\cosh^2 \frac{\phi}{\sqrt{3}}}\;,
\end{equation}

one finds that equation (\ref{ein4}) reduces to (\ref{ein2}) which are exactly the Einstein equations for EMCS theory described by action $\mathcal{S}_2$.

Now let's make an examination on whether the new solution is consistent with all the equations of motion. The equations of motion for the Maxwell field is
\begin{equation}
\partial \mu \left[\sqrt{-\bar{g}} K_{1}(\phi) \bar{F}^{\mu \nu}+\sqrt{-\bar{g}} L_{1}(\phi) {}^{*} \bar {{F}} ^{\mu \nu}\right]=0\;.
\end{equation}
It's not hard to prove that the quantity in the square bracket remains unchanged under conformal translations. So we have
\begin{equation}
\partial \mu\left[\sqrt{{-\tilde{g}}} K_{2}(\psi) {\tilde{F}}^{\mu \nu}+\sqrt{{-\tilde{g}}} L_{2}(\psi){}^{*}{\tilde{F}}^{\mu \nu}\right]=0\;,
\end{equation}
which is the equation of motion for Maxwell field in action $\mathcal{S}_2$.

The last step is to check the scalar field equations in  $\mathcal{S}_1$ and $\mathcal{S}_2$ which are
\begin{equation}
\label{eqphi}
4 \bar{\nabla}_{\mu} \bar{\nabla}^{\mu} \phi-K_{1}^{\prime}(\phi) \bar{F}^{2}-L_{1}^{\prime}(\phi)\bar{F}_{\mu\nu} {}^{*} \bar {{F}} ^{\mu\nu}=0\;,
\end{equation}
and
\begin{equation}
\label{eqpsi1}
4 {\tilde{\nabla}}_{\mu} {\tilde{\nabla}}^{\mu} \psi-\frac{2\psi \tilde{R}}{3}-K_{2}^{\prime}(\psi) {\tilde{F}}^{2}-L_{2}^{\prime}(\psi) {\tilde{F}}_{\mu\nu} {}^{*}  {{\tilde{F}}} ^{\mu\nu}=0\;,
\end{equation}
respectively. What we are going to check is that, basing on the equations of motion which have been previously proved together with the scalar field equation  (\ref{eqphi}), the equation (\ref{eqpsi1}) is automatically satisfied. Indeed, the trace of Einstein equation (\ref{ein2}) gives $-\tilde{R} \Omega^2=2\psi \tilde{\nabla}_{\mu}\tilde{\nabla}^{\mu}\psi$. Hence the left hand of side for equation (\ref{eqpsi1}) turns out to be
\begin{equation}
\begin{aligned}
\label{eqpsi2}
4 {\tilde{\nabla}}_{\mu} {\tilde{\nabla}}^{\mu} \psi-\frac{2\psi \tilde{R}}{3}-K_{2}^{\prime}(\psi) {\tilde{F}}^{2}-L_{2}^{\prime}(\psi) {\tilde{F}}_{\mu\nu} {}^{*}  {{\tilde{F}}} ^{\mu\nu}=&4 {\nabla}_{\mu} {\nabla}^{\mu} \psi+\frac{2\psi}{3}\left(\frac{2\psi \tilde{\nabla}_{\mu}\tilde{\nabla}^{\mu}\psi}{\Omega^2}\right)-K_2^{\prime}(\psi)\tilde{F}^2-L_{2}^{\prime}(\psi) {\tilde{F}}_{\mu\nu} {}^{*} {{\tilde{F}}} ^{\mu\nu}\\
&=4 {\tilde{\nabla}}_{\mu} {\tilde{\nabla}}^{\mu} \psi\left(1+\frac{\psi^2}{3\Omega^2}\right)-K_2^{\prime}(\psi)\tilde{F}^2-L_{2}^{\prime}(\psi) {\tilde{F}}_{\mu\nu} {}^{*} {{\tilde{F}}} ^{\mu\nu}\\
&=\frac{4 {\tilde{\nabla}}_{\mu} {\tilde{\nabla}}^{\mu} \psi}{\Omega^2}-K_2^{\prime}(\psi)\tilde{F}^2-L_{2}^{\prime}(\psi) {\tilde{F}}_{\mu\nu} {}^{*} {{\tilde{F}}} ^{\mu\nu}\;.
\end{aligned}
\end{equation}
On the other hand, we have
\begin{equation}
\label{eqphileft1}
\begin{aligned}
 \bar{\nabla}_{\mu} \bar{\nabla}^{\mu} \phi=\frac{1}{\sqrt{-\bar{g}}} \left(\sqrt{-\bar{g}}\bar{g}^{\beta\alpha}\phi,_{ \alpha}\right),_{\beta}=\frac{1}{\Omega^4 \sqrt{-\tilde{g}}}\left(\sqrt{-{\tilde{g}}}{\tilde{g}}^{\beta\alpha}\psi,_{ \alpha}\right),_{\beta}=\frac{1}{\Omega^4} {\tilde{\nabla}}_{\mu} {\tilde{\nabla}}^{\mu} \psi\;,
\end{aligned}
\end{equation}
and
\begin{equation}
\label{eqphileft2}
\begin{aligned}
K_{1}^{\prime}(\phi)\bar{F}^2=\frac{d K_1(\phi)}{d \phi} \bar{F}^2=\frac{d K_2(\psi)}{d\psi} \frac{d \psi}{d \phi} \frac{F^2}{\Omega^4}=\frac{d K_2(\psi)}{d\psi} \frac{\tilde{F}^2}{\Omega^4}=K_2^{\prime}(\psi)\frac{\tilde{F}^2}{\Omega^2}\;,\ \ \ \ L_{1}^{\prime}(\phi)\bar{F}_{\mu\nu}  {}^{*} {\bar{F}} ^{\mu\nu}=L_{2}^{\prime}(\psi) \frac{{\tilde{F}}_{\mu\nu} {}^{*} {{\tilde{F}}} ^{\mu\nu}}{\Omega^2}\;.
\end{aligned}
\end{equation}
Substituting equations (\ref{eqphileft1}) and (\ref{eqphileft2}) into equation (\ref{eqphi}), we obtain
\begin{equation}
\frac{4 {\tilde{\nabla}}_{\mu} {\tilde{\nabla}}^{\mu} \psi}{\Omega^2}-K_2^{\prime}(\psi)\tilde{F}^2-L_{2}^{\prime}(\psi) {\tilde{F}}_{\mu\nu} {}^{*} {{\tilde{F}}} ^{\mu\nu}=0\;,
\end{equation}
 which is exactly the last line of equation (\ref{eqpsi2}). Up to this point, equation (\ref{eqpsi1}) is satisfied.  Q.E.D.

We note that in the case of $K_{1}(\phi)=1$ and $L_{1}(\phi)=0$, the theorem above reduces to  Bekenstein's theorem. In the next two sections, as applications of our theorem, we shall take the static dilaton black hole solution  and the rotating dilation black hole solution as seed solutions to generate two new solutions. As is known to all, it is usually very hard to obtain rotating black hole solution in General Relativity. However, by using this theorem, one can straightforwardly get rotating black hole solution in conformal dilaton theory via conformal transformations.

 \section{generating new static DILATON BLACK HOLE SOLUTION}
The action of Einstein-Maxwell-dilaton theory is given by
\begin{equation}
\mathcal{S}_1=\int d^{4} x \sqrt{-\bar{g}}\left[\bar{R}-2(\bar{\nabla} \phi)^{2}-e^{-2 \alpha \phi} \bar{F}^{2}\right]\;.
\end{equation}
Now we apply our theorem to this theory which has $K_{1}(\phi)=e^{-2\alpha \phi}$ and $L_{1}(\phi)=0$. The corresponding spherically symmetric black hole solution takes the form \cite{horne1992rotating}
\begin{equation}
\label{metric1}
d \bar{s}^{2}=-\lambda^{2} d t^{2}+\frac{d r^{2}}{\lambda^{2}}+R^{2} d\Omega_2^2\;,
\end{equation}
\begin{equation}
e^{2 \phi}=\left(1-\frac{r_{-}}{r}\right)^{\frac{2 \alpha}{1+\alpha^{2}}}\;,\ \ \ \  \bar{F}_{t r}=\frac{\bar{Q}}{r^{2}}\;,
\end{equation}
with
\begin{equation}
\lambda^{2}=\left(1-\frac{r_{+}}{r}\right)\left(1-\frac{r_{-}}{r}\right)^{\frac{1-\alpha^{2}}{1+\alpha^{2}}}\;,\ \ \ \ \ R=r\left(1-\frac{r_{-}}{r}\right)^{\frac{\alpha^{2}}{1+\alpha^{2}}}\;.
\end{equation}

The physical mass $\bar{M}$, charge $\bar{Q}$ are related to $r_{-}$ and $r_{+}$ as follows

\begin{equation}
\bar{M}=\frac{r_{+}}{2}+\left(\frac{1-\alpha^{2}}{1+\alpha^{2}}\right) \frac{r_{-}}{2}\;, \ \ \ \
\bar{Q}=\left(\frac{r_{+} r_{-}}{1+\alpha^{2}}\right)^{\frac{1}{2}}\;.
\end{equation}

According to the theorem, the scalar field $\psi$ in corresponding $\mathcal{S}_2$ is
\begin{equation}
\psi=\sqrt{3}\tanh \frac{\phi}{\sqrt{3}}=\sqrt{3} \frac{\left(1-\frac{r_{-}}{r}\right)^{\frac{2 \alpha}{\sqrt{3}(1+\alpha^{2})}}-1}{\left(1-\frac{r_{-}}{r}\right)^{\frac{2 \alpha}{\sqrt{3}(1+\alpha^{2})}}+1}\;,
\end{equation}
while the Maxwell field remains unchanged.  So we have
\begin{equation}
K_{2}(\psi)=e^{-2\alpha \sqrt{3}\text{arctanh} (\psi/\sqrt{3})}=\left(\frac{3+\sqrt{3}\psi}{3-\sqrt{3}\psi}\right)^{-\sqrt{3}\alpha}\;,\ \ \ \ L_{2}(\psi)=0\;,
\end{equation}
and
\begin{equation}
\begin{aligned}
\Omega^2=&1-\frac{\psi^2}{3}=\left[\frac{2}{\left(1-\frac{r_{-}}{r}\right)^{\frac{\alpha}{\sqrt{3}\left(1+\alpha^{2}\right)}}+\left(1-\frac{r-}{r}\right)^{\frac{-\alpha}{\sqrt{3}\left(1+\alpha^{2}
\right)}}}\right]^{2}\;.
\end{aligned}
\end{equation}
$\Omega$ approaches unit one when $r$ approaches $+\infty$. The black hole solution corresponding to $\mathcal{S}_{2}$ is then
\begin{equation}
\label{ele2}
d\tilde{s}^2=\frac{1}{\Omega^2}\left(-\lambda^{2} d t^{2}+\frac{d r^{2}}{\lambda^{2}}+R^{2} d\Omega_2^2\right)\;.
\end{equation}
Since we have $\Omega^2 \to 1$ in the spacial infinity, the spacetime is also asymptotically flat.

Now we turn to the  thermodynamics of the old (equation (\ref{metric1})) and new black hole (equation (\ref{ele2})) solutions which are connected by conformal transformation. The quantities corresponding to the old and new black holes are labeled with bar and tilde, respectively. For the new black hole, the energy can be calculated by Komar Mass. If one denotes the timelike vector by $k$ which is normalized at spatial infinity, the Komar mass is then given by
 \begin{equation}\label{km}
 M_{\mathrm{K}}:=-\frac{1}{8 \pi} \oint_{\mathscr{S}_{t}} \nabla^{\mu} k^{v} \mathrm{~d} S_{\mu v}\;.
 \end{equation}
 The integration is integrated on a closed 2-surface $\mathscr{S}_{t}$ in $\Sigma_{t}$. Here $\Sigma_{t}$ provides a $3+1$ foliation $(\Sigma_{t})_{t\in \mathbb{R}}$ of the spacetime. The line element of the surface is
 \begin{equation}
 	\label{2surface}
 	\mathrm{d} S_{\mu \nu}=\left(s_{\mu} n_{v}-n_{\mu} s_{v}\right) \sqrt{q} \mathrm{~d}^{2} y\;,
 \end{equation}
 where $n$ is the unit timelike vector normal to $\Sigma_{t}$. $s$ is the unit vector normal to $\mathscr{S}_{t}$ within $\Sigma_{t}$ which is oriented towards the exterior of $\mathscr{S}_{t}$. $y^a=(y^1,y^2)$ are coordinates spanning  $\mathscr{S}_{t}$. $q$ is defined as $q:=det(q_{ab})$ and $q_{ab}$ are the components of the metric induced by $g$ on  $\mathscr{S}_{t}$.  In order to calculate of the Komar mass (and the angular momentum in the next sections), we closely follow the detailed procedures given by Gourgloulhon\cite{gourgoulhon20123+}.

The timelike  vector $\partial_{t}=\boldsymbol{k}$ of the spacetime should be decomposed into lapse and shift vector
\begin{equation}
\boldsymbol{k}=N \boldsymbol{n}+\boldsymbol{\beta}\;.
\end{equation}
Then the Komar mass becomes the form
\begin{equation}
\label{mass1}
M_{\mathrm{K}}=\frac{1}{4 \pi} \oint_{\mathscr{S}_{t}}\left(s^{i} D_{i} N-K_{i j} s^{i} \beta^{j}\right) \sqrt{q} \mathrm{~d}^{2} y\;.
\end{equation}
The shift vector vanishes for a static spacetime, thus, only the first term in the integrand makes contribution to the integration.
The unit normal vector $s^i$ to $\mathscr{S}_{t}$ is
\begin{equation}
s^i=\left(\sqrt{1/g_{11}},0,0\right)\;,
\end{equation}
and $N={\sqrt{-1/g^{00}}}$, $\sqrt{q}=\sqrt{g_{22}g_{33}}$. Therefore, we obtain
\begin{equation}
\tilde{s}^1=\sqrt{\Omega^2 \lambda^2}\sim 1+O(1/r)\;,
\end{equation}
\begin{equation}
\partial_{r}\tilde{N}=\partial_{r}\sqrt{\frac{\lambda^2}{\Omega^2}} \sim \left[\frac{r_{+}}{2}+\left(\frac{1-\alpha^{2}}{1+\alpha^{2}}\right)\frac{r_{-}}{2}\right]/r^2+O(1/r^3)\;,
\end{equation}
\begin{equation}
\sqrt{\tilde{q}}=R^2/\Omega^2\sin\theta\sim r^2\sin\theta+O(r)\;.
\end{equation}
The Komar mass is
\begin{equation}
\tilde{M}=\frac{1}{4 \pi} \oint_{\mathscr{S}_{t}}\left[\frac{r_{+}}{2}+\left(\frac{1-\alpha^{2}}{1+\alpha^{2}}\right)\frac{r_{-}}{2}\right]\sin\theta d\theta d\varphi=\frac{r_{+}}{2}+\left(\frac{1-\alpha^{2}}{1+\alpha^{2}}\right)\frac{r_{-}}{2}=\bar{M}\;,
\end{equation}
namely the mass is invariant.

The electric potential $\bar{\Phi}$ conjugated to $\bar{Q}$ is given by $\bar{\Phi}=\bar{A}_{0}(r_{+})=\frac{\bar{Q}}{r_{+}}$.
The temperature $\bar{T}$ is
\begin{equation}
\bar{T}=\frac{1}{4\pi}(\lambda^2)'|_{r_{+}}=\frac{(1-r_{-}/r_{+})^{\frac{1-\alpha^2}{1+\alpha^2}}}{4\pi r_{+}}\;.
\end{equation}
The conjugate variable of temperature can be regraded as the Wald entropy \cite{wald1993black,iyer1994some}.
Given the action $I$, the Wald entropy is
\begin{equation}
\label{entropydef}
S=2 \pi \int_{\mathscr{S}_{t}} \frac{\delta I}{\delta R_{\mu \nu \alpha \beta}} \epsilon_{\mu \alpha} \epsilon_{\nu \beta} \sqrt{h} d\Omega_2\;,
\end{equation}
where $\epsilon^{\mu \nu}$ is the binormal to the horizon, $h$ the induced metric on  $\mathscr{S}_{t}$. The
variation of the action with respect to $R_{\mu \nu \alpha \beta}$ is to be carried out by regarding the Riemann
tensor $R_{\mu \nu \alpha \beta}$ as formally independent on the metric $g_{\mu\nu}$.
In the static spacetime, the event horizon is $\mathscr{S}_{t}$ which has two normal directions along $r$
and $t$. We can construct an antisymmetric 2-tensor $\epsilon_{\mu\nu}$ along these directions so that
$\epsilon_{rt}=\epsilon_{tr}=-1$.

For the action $\mathcal{S}_{1}$ we only need to concentrate on the scalar curvature
\begin{equation}
\mathcal{L}=\frac{1}{16 \pi} \bar{R}_{\mu \nu \alpha \beta} \bar{g}^{\nu \alpha} \bar{g}^{\mu \beta}\;,\ \ \  \quad \frac{\partial \mathcal{L}}{\partial
\bar{R}_{\mu \nu \alpha \beta}}=\frac{1}{16 \pi} \frac{1}{2}\left(\bar{g}^{\mu \alpha} \tilde{g}^{\nu \beta}-\bar{g}^{\nu \alpha} \bar{g}^{\mu \beta}\right)\;.
\end{equation}
Then the Wald entropy yields
\begin{equation}
\begin{aligned}
\bar{S} &=\frac{1}{8} \int \frac{1}{2}\left(\bar{g}^{\mu \alpha} \bar{g}^{\nu \beta}-\bar{g}^{\nu \alpha} \bar{g}^{\mu \beta}\right)\left(\bar{\epsilon}_{\mu \nu} \bar{\epsilon}_{\alpha \beta}\right) \sqrt{\bar{h}} d\Omega_2 \\
&=\frac{1}{4} \int_{\mathscr{S}_{t}} \sqrt{\bar{h}} d\Omega_2=\frac{\bar{A}_{H}}{4}=r^2_{+}\left(1-\frac{r_{-}}{r_{+}}\right)^{\frac{2\alpha^{2}}{1+\alpha^{2}}}\;,
\end{aligned}
\end{equation}
which is equal to the Bekenstein-Hawking entropy \cite{bekenstein2020black} as expected. Armed with these expressions, one can construct the first law of thermodynamics and the Smarr formula
\begin{equation}
d\bar{M}= \bar{T} d\bar{S} +\bar{\Phi} d\bar{Q}\;,\ \ \ \ \ \bar{M}=2\bar{T}\bar{S}+\bar{\Phi}\bar{Q}\;.
\end{equation}

We have discussed the thermodynamics of the dilaton black hole in the preceding contents, and now we will discuss the conformal version $\mathcal{S}_2$. the temperature of the black hole (\ref{ele2}) is
\begin{equation}
\tilde{T}=\frac{1}{4 \pi} \frac{\left(\frac{\lambda^{2}}{\Omega^{2}}\right)^{\prime}}{\sqrt{\frac{\lambda^{2}}{\Omega^{2}} \cdot \frac{1}{\Omega^{2} \lambda^{2}}}}|_{r_{+}}=\frac{1}{4 \pi} \frac{\frac{\left(\lambda^{2}\right)^{\prime}}{\Omega^{2}}+\left(\frac{1}{\Omega^{2}}\right)^{\prime} \lambda^{2}}{\frac{1}{\Omega^{2}}}|_{r_{+}}=\frac{1}{4\pi}(\lambda^2)'|_{r_{+}}=\bar{T}\;.
\end{equation}
The penultimate equation holds because $\lambda^2 |_{r_{+}}=0$. Thus the temperature remains unchanged after conformal transformation.
 Its conjugate variable, entropy now equals
 \begin{equation}
\tilde{S}=\frac{1-\frac{1}{3}\psi^2}{4} \int_{\mathscr{S}_{t}} \sqrt{\tilde{h}} d\Omega_2=\frac{ \Omega^2 \tilde{A}_H }{4}=r^2_{+}\left(1-\frac{r_{-}}{r_{+}}\right)^{\frac{2\alpha^{2}}{1+\alpha^{2}}}=\bar{S}\;.
 \end{equation}
 Namely, the Wald temperature remains unchanged. We should emphasize that the entropy here is not the Bekenstein-Hawking entropy $\tilde{A}_H/4$, but the Wald entropy $\Omega^2\tilde{A}_H/4$. Why? This argument may be understood in terms of higher curvature gravity, such as Gauss-Bonnet gravity \cite{jacobson1995black,jacobson1995increase}. It is found \cite{cai2002gauss,clunan2004gauss} that, for Gauss-Bonnet black holes in AdS spaces, it is only the Wald entropy which plays the role Noether charge that can be applied in the first law of thermodynamics. The reason for this point is that in the action, the geometry quantities making contribution to the Wald entropy include not only the scalar curvature but also the Gauss-Bonnet term. 
 
 For non-minimally coupled scalar fields which are considered here, Ashtekar suggested that in the theory governed by 
 \begin{equation}
 \mathcal{S}\left[g_{a b}, \phi\right]=\int \mathrm{d}^{4} x \sqrt{-g}\left[\frac{1}{16 \pi G} f(\phi) R-\frac{1}{2} g^{a b} \partial_{a} \phi \partial_{b} \phi-U(\phi)\right]\;,
 \end{equation}
 
 where $R$ is the scalar curvature of the metric $g_{ab}$ and $U$ is a potential for the scalar field, the Wald
 entropy is given by\cite{ashtekar2003non}
 \begin{equation}
 S=\frac{1}{4 \ell_{\mathrm{Pl}}^{2}} \oint_{\mathscr{S}_{t}} f(\phi) \mathrm{d}^{2} V\;.
 \end{equation}

 Observing action $\mathcal{S}_{2}$, we find it is not $\tilde{R}$ but $(1-\frac{\psi^2}{3})\tilde{R}$ which contributes the Wald entropy. Therefore, we should replace Bekenstein-Hawking entropy with Wald entropy for the new black holes, which equals to  the hawking entropy multiplied by $\Omega^2$.
 For more instances of theories involving a conformally linked scalar field, one can refer to \cite{winstanley2004classical,barlow2005thermodynamics}.

 Since the Faraday tensor is invariant after conformal transformation $\tilde{F}_{\mu\nu}=\bar{F}_{\mu\nu}$, we have $\tilde{A}_{\mu}=\bar{A}_{\mu}$. Therefore the conjugate variable to charge $\tilde{Q}$, $\tilde{\Phi}$ is equal to $\bar{\Phi}$.
The electric charge $\tilde{Q}$ is determined by
\begin{equation}
\tilde{Q}\equiv \frac{1}{4 \pi} \int_{\mathscr{S}_{t}} K_{2}(\psi)^{*} F d\Omega_2\;.
\end{equation}
Note that $\bar{Q}\equiv \frac{1}{4 \pi} \int_{S^{2}} K_{1}(\phi)^{*} F d \Omega_2$ and $K_1(\phi)=K_2(\psi)$, thus we have $\bar{Q}=\tilde{Q}$.

In summary, both the old and the new black holes have the same thermodynamics. Concretely, they have same Komar masses $\tilde{M}=\bar{M}$, same electric charges $\tilde{Q}=\bar{Q}$, same electric potentials $\tilde{\Phi}=\bar{\Phi}$, same temperatures $\tilde{T}=\bar{T}$ and same Wald entropies $\tilde{S}=\bar{S}$ (but different Bekenstein-Hawking entropies). We emphasize that although they have same physical quantities in the first law of thermodynamics  $dM=TdS+\Phi dQ $ , in essence, they are completely different spacetimes.

\section{generating new ROTATING DILATON BLACK HOLE}
A rotating dilaton black hole solution with a special coupling constant $\alpha=\sqrt{3}$ is found by Frolov et al. \cite{frolov1987charged}. For this value of $\alpha$, the action is simply the Kaluza-Klein action which is obtained by dimensionally reduction of five dimensional vacuum Einstein action. Applying our theorem, we can obtain a new rotating black hole solution.

The metric for a rotating dilaton black hole with $\alpha=\sqrt{3}$ is given by \cite{horne1992rotating}

\begin{equation}
\begin{aligned}
\label{rometric}
d \bar{s}^{2}=-\frac{1-Z}{B} d t^{2}-\frac{2 a Z \sin ^{2} \theta}{B \sqrt{1-v^{2}}} d t d \phi
+\left[B\left(r^{2}+a^{2}\right)+a^{2} \sin ^{2} \theta \frac{Z}{B}\right] \sin ^{2} \theta d \phi^{2}+B \frac{\Sigma}{\Delta_{0}} d r^{2}+B \Sigma d \theta^{2}\;,
\end{aligned}
\end{equation}
where
\begin{equation}
B=\sqrt{1+\frac{v^{2} Z}{1-v^{2}}}, \quad Z=\frac{2 m r}{\Sigma}, \quad \Delta_{0}=r^{2}+a^{2}-2 m r,\quad \Sigma=r^{2}+a^{2} \cos ^{2} \theta\;.
\end{equation}
Here $m$ and $a$ are related to the mass and angular momentum of the black hole, respectively. The constant $v$ is the velocity of the boost.
The vector potential and the dilaton field are
\begin{equation}
A_{t}=\frac{v}{2\left(1-v^{2}\right)} \frac{Z}{B^{2}}\;,\ \ \  \quad A_{\phi}=-a \sin ^{2} \theta \frac{v}{2 \sqrt{1-v^{2}}} \frac{Z}{B^{2}}\;,\ \ \
\phi=-\frac{\sqrt{3}}{2} \log B\;.
\end{equation}
The physical mass $\bar{M}$, charge $\bar{Q}$ and angular momentum $\bar{J}$ are given by
\begin{eqnarray}
\bar{M}=m\left[1+\frac{v^{2}}{2\left(1-v^{2}\right)}\right]\;,\ \ \ \bar{Q}=\frac{m v}{1-v^{2}}\;,\ \ \ \ \bar{J}=\frac{m a}{\sqrt{1-v^{2}}}\;.
\end{eqnarray}

The theorem tells us $K_{2}(\psi)=\left(\frac{3+\sqrt{3}\psi}{3-\sqrt{3}\psi}\right)^{-3}$ and the conformal factor is
\begin{equation}
\begin{aligned}
\Omega^2=\left(\frac{2}{e^{\phi/\sqrt{3}}+e^{-\phi/\sqrt{3}}}\right)^2=\left(\frac{2}{\frac{1}{\sqrt{B}}+\sqrt{B}}\right)^2\;.
\end{aligned}
\end{equation}
So the line element for the black hole gives
\begin{equation}
\begin{aligned}
\label{rometric2}
d\tilde{s}^2=\frac{1}{\Omega^2}\left\{-\frac{1-Z}{B} d t^{2}-\frac{2 a Z \sin ^{2} \theta}{B \sqrt{1-v^{2}}} d t d \phi
+\left[B\left(r^{2}+a^{2}\right)+a^{2} \sin ^{2} \theta \frac{Z}{B}\right] \sin ^{2} \theta d \phi^{2}+B \frac{\Sigma}{\Delta_{0}} d r^{2}+B \Sigma d \theta^{2} \right\}\;.
\end{aligned}
\end{equation}

The scalar field is
\begin{equation}
\psi=\sqrt{3}\tanh \frac{\phi}{\sqrt{3}}=\sqrt{3} \frac{e^{\frac{2\phi}{\sqrt{3}}}-1}{e^{\frac{2\phi}{\sqrt{3}}}+1}=\sqrt{3} \frac{1-B}{1+B}\;.
\end{equation}

To compute the Komar mass $\tilde{M}$ of rotating black hole, one should consider the extrinsic curvature term in (\ref{mass1}). It is obvious the components of shift vector $\beta^{i}=g^{0i}/g^{00}$ are conformal invariant. Taken into account the ``$3+1$'' foliation associated with the standard Boyer Lindquist coordinates $(t,r,\theta,\varphi)$, the non-vanishing component of $\beta^{i}$ is $\beta^{\varphi}$.
Therefore, we can only concentrate on $K_{r\varphi}s^{r}\beta^{\varphi}$. The extrinsic curvature $K_{r\varphi}$ is evaluated via $2 N K_{i j}=\mathscr{L}_{\beta} \gamma_{i j}$ in the case of $\partial \gamma_{i j} / \partial t=0$, with $\gamma_{i j}$ the induced metric on the space.
Then we have

\begin{equation}
\label{extrinsic}
K_{r \varphi}=\frac{1}{2 N} \mathscr{L}_{\beta} \gamma_{r \varphi}=\frac{1}{2 N}(\beta^{\varphi} \underbrace{\frac{\partial \gamma_{r \varphi}}{\partial \varphi}}_{0}+\gamma_{\varphi \varphi} \frac{\partial \beta^{\varphi}}{\partial r}+\gamma_{r \varphi} \underbrace{\frac{\partial \beta^{\varphi}}{\partial \varphi}}_{0})=\frac{1}{2 N} \gamma_{\varphi \varphi} \frac{\partial \beta^{\varphi}}{\partial r}\;.
\end{equation}

Hence,
\begin{equation}
\label{mass2}
K_{r \varphi}s^r\beta^{\varphi}\sqrt{q}=\frac{s^r}{2 N} \gamma_{\varphi \varphi} \frac{\partial \beta^{\varphi}}{\partial r}\beta^{\varphi}\sqrt{q}=\frac{1}{4}\sqrt{\frac{-g^{00}}{g_{11}}}g_{33}\sqrt{g_{22}g_{33}}\frac{\partial(\frac{g^{03}}{g^{00}})^2}{\partial r}\;.
\end{equation}
In the limit of $r \to \infty$, we get $\sqrt{\frac{-\tilde{g}^{00}}{\tilde{g}_{11}}} \sim 1$, $\tilde{g}_{33} \sim r^2 \sin\theta$, $\sqrt{\tilde{g}_{22}\tilde{g}_{33}} \sim r^2 \sin\theta$, $\frac{\partial(\frac{\tilde{g}^{03}}{\tilde{g}^{00}})^2}{\partial r} \sim \frac{24a^2m^2}{(-1+v^2)r^7} $. Therefore, we have $\tilde{K}_{r \varphi}\tilde{s}^r\tilde{\beta}^{\varphi}\sqrt{\tilde{q}} \sim O(1/r^3)$ which does not contribute to the integral in definition of Komar mass.

Then considering the term $\tilde{s^i}D_{i}\tilde{N}\sqrt{\tilde{q}}$ again, we have
\begin{equation}
\tilde{s}^r \sim 1 \;,\ \ \ \ \partial_{r}\tilde{N} \sim \frac{m(2-v^2)}{2(1-v^2)r^2}\;,\ \ \ \ \ \sqrt{\tilde{q}} \sim r^2 \sin\theta\;.
\end{equation}

As a result, the Komar mass for the rotating black hole (\ref{rometric2}) according to equation (\ref{mass1}) is
\begin{equation}
\tilde{M}=\frac{1}{4 \pi} \oint_{\mathscr{S}_{t}}\frac{m(2-v^2)}{2(1-v^2)}\sin\theta d\theta d\varphi=\frac{m(2-v^2)}{2(1-v^2)}=\bar{M}\;.
\end{equation}

As for the electric charge $\tilde{Q}$ and the Wald entropy $\tilde{S}$ , they also remain unchanged after the conformal transformation, i.e. $\tilde{Q}=\bar{Q}$, $\tilde{S}=\bar{S}$.  To calculate the temperature of this rotating black hole, one can refer to \cite{ma2008hawking} where it shows the temperature is given by
\begin{equation}\label{tem}
T_{H}=\lim _{r \rightarrow r_{H}} \frac{\partial_{r} \sqrt{-g_{tt}-2 g_{t\phi}\Omega _{H}-g_{\phi\phi}\Omega _{H}^{2}}}{2 \pi \sqrt{g_{r r}}}\;,
\end{equation}
for the  four-dimensional rotating black hole with the metric
\begin{equation}
\begin{aligned}
d s^{2}=&g_{t t}(r, \theta) d t^{2}+g_{r r}(r, \theta) d r^{2}+g_{\theta \theta}(r, \theta) d \theta^{2} \\
&+g_{\phi \phi}(r, \theta) d \phi^{2}+2 g_{t \phi}(r, \theta) d t d \phi\;.
\end{aligned}
\end{equation}
Here $\Omega_{H}=-\frac{g_{tt}}{g_{t\phi}}|_{r\to r_{H}}$ is  the conjugate variable to angular momentum of the black hole- angular velocity of the black hole. It is very obvious the angular velocity of black hole  is not changed under conformal transformation $g_{\mu\nu} \to \frac{1}{\Omega^2}g_{\mu\nu}$.
If we denote $ \sqrt{-g_{tt}-2 g_{t\phi}\Omega _{H}-g_{\phi\phi}\Omega _{H}^{2}}$ by $X$, then we get $X=0$ on the horizon since  the horizon is a null surface. Then, under conformal transformation $g_{\mu\nu} \to \frac{1}{\Omega^2}g_{\mu\nu}$, we obtain

\begin{equation}
\frac{\partial r \sqrt{X}}{2 \pi \sqrt{g_{r r}}}\to \frac{\partial r \sqrt{\frac{X}{\Omega^{2}}}}{2 \pi \sqrt{\frac{g_{r r}}{\Omega^{2}}}}=\frac{\partial r \sqrt{X}}{2 \pi \sqrt{g_{r r}}}+\frac{\partial r\left(\frac{1}{\Omega^{2}}\right) X}{2 \pi \sqrt{\frac{g_{r r}}{\Omega^{2}}}}=	\frac{\partial r \sqrt{X}}{2 \pi \sqrt{g_{r r}}}\;.
\end{equation}

Then it follows that $\bar{T}=\tilde{T}$ from (\ref{tem}). The electric potential is defined by
\begin{equation}
\Phi=\left.A_{\mu} \chi^{\mu}\right|_{r \rightarrow \infty}-\left.A_{\mu} \chi^{\mu}\right|_{r=r_{H}}\;,
\end{equation}
where  $\chi=\partial_{t}+\Omega_{H} \partial_{\phi}$ is the null generator of the horizon and $A_{\mu}$ denotes the vector potential.
Since $\bar{\Omega}_{H}=\tilde{\Omega}_{H}$ we conclude $\bar{\chi}_{\mu}=\tilde{\chi}_{\mu}$. Taking into account $\bar{A}_{\mu}=\tilde{A}_{\mu}$, we obtain  $\bar{\Phi}=\tilde{\Phi}$.

Turning to angular momentum for (\ref{rometric2}), we have the definition of Komar angular momentum \cite{komar1959covariant,gourgoulhon20123+}

\begin{equation}
\label{angularmomentum}
J_{\mathrm{K}}:=\frac{1}{16 \pi} \oint_{\mathscr{S}_{t}} \nabla^{\mu} \varphi^{\nu} \mathrm{d} S_{\mu \nu}\;,
\end{equation}
 where $\varphi^{\nu}$ is the Killing vector, and $\mathscr{S}_{t}$ share the same meaning as  in the definition of Komar mass in section III. It is natural to choose a foliation adapted to the axisymmetric in the sense that the Killing
vector $\varphi$ is tangent to the hypersurface $\Sigma_{t}$. Then we have $n*\varphi=0$ and the integrand in (\ref{angularmomentum}) is

\begin{equation}
\begin{aligned}
\nabla^{\mu} \varphi^{v} \mathrm{~d} S_{\mu v} &=\nabla_{\mu} \varphi_{v}\left(s^{\mu} n^{v}-n^{\mu} s^{v}\right) \sqrt{q} \mathrm{~d}^{2} y=2 \nabla_{\mu} \varphi_{v} s^{\mu} n^{v} \sqrt{q} \mathrm{~d}^{2} y \\
&=-2 s^{\mu} \varphi_{v} \nabla_{\mu} n^{v} \sqrt{q} \mathrm{~d}^{2} y=2 K_{i j} s^{i} \varphi^{j} \sqrt{q} \mathrm{~d}^{2} y\;.
\end{aligned}
\end{equation}
So Eq.(\ref{angularmomentum}) becomes

\begin{equation}
\label{angularmomentum2}
J_{\mathrm{K}}=\frac{1}{8 \pi} \oint_{\mathscr{S}_{t}} K_{i j} s^{i} \varphi^{j} \sqrt{q} \mathrm{~d}^{2} y\;.
\end{equation}

Let us use the ``$3+1$'' foliation associated with the standard Boyer-Lindquist coordinates $(t,r,\theta,\phi)$ and evaluate the integral (\ref{angularmomentum2}) by choosing for $\mathcal{S}_{t}$ as a sphere $r=const$. Then, we have $y^a=(\theta,\phi)$.  The Boyer-Lindquist components of $\varphi$ are $\varphi^i=(0,0,1)$ and those of $s$ are $s^i=(s^r,0,0)$ because $\gamma_{i,j}$ is diagonal in this coordinate system. Thus (\ref{angularmomentum2}) reduces to
\begin{equation}
J_{\mathrm{K}}=\frac{1}{8 \pi} \oint_{r=\mathrm{const}} K_{r \varphi} s^{r} \sqrt{q} \mathrm{~d} \theta \mathrm{d} \varphi\;.
\end{equation}
The extrinsic curvature $K_{r\varphi}$ can be evaluated via $2 N K_{i j}=\mathscr{L}_{\beta} \gamma_{i j}$ when we have $\partial \gamma_{i j} / \partial t=0$
and the timelike vector $\partial_{t}$ can be decomposed into $\partial_{t}=N \boldsymbol{n}+\boldsymbol{\beta}$. One can read off the components of the shift vector $\beta^{i}=g^{0i}/g^{00}$ for the rotating conformal dilaton black hole

\begin{equation}
\left(\tilde{\beta}^{r}, \tilde{\beta}^{\theta}, \tilde{\beta}^{\varphi}\right)=\left(0,0, \frac{-a z}{B \sqrt{1-v^{2}}\left[B\left(r^{2}+a^{2}\right)+a^{2} \sin ^{2} \theta \frac{z}{B}\right]}\right)\;.
\end{equation}

Keeping in mind the expression for extrinsic curvature (\ref{extrinsic}), we find the Komar angular momentum is
\begin{equation}
\label{Jdef1}
J_{\mathrm{K}}=\frac{1}{16 \pi} \oint_{r=\mathrm{const}} \frac{s^{r}}{N} \gamma_{\varphi \varphi} \frac{\partial \beta^{\varphi}}{\partial r} \sqrt{q} \mathrm{~d} \theta \mathrm{d} \varphi\;.
\end{equation}
For the new rotating dilaton black hole, we have
 \begin{equation}
 \label{intj1}
\frac{\tilde{s}^{r}}{\tilde{N}} \tilde{\gamma}_{\varphi \varphi} \sqrt{\tilde{q}}=\sqrt{\frac{-\tilde{g}^{00}}{\tilde{g}_{11}}} \tilde{g}_{33} \sqrt{\tilde{g}_{22}\tilde{g}_{33}} =\frac{1}{\Omega^{2}} \frac{\bar{s}^{r}}{\bar{N}} \bar{\gamma}_{\varphi \varphi} \sqrt{\bar{q}}\;,\ \ \ \ \
\tilde{\beta}^{i} =\tilde{g}^{0i}/\tilde{g}^{00}= \bar{\beta}^{i}\;.
\end{equation}
In the limit of $r \to \infty$, we have $\bar{s}^r \sim 1+O(1/r)$, $\frac{1}{\bar{N}} \sim 1+O(1/r)$, $\bar{\gamma}_{\varphi \varphi} \sim r^2\sin^2\theta+O(r)$, ${\sqrt{\bar{q}}} \sim r^2 \sin\theta+O(r)$, $ \frac{\partial \bar{\beta}^{\varphi}}{\partial r} \sim \frac{6am}{\sqrt{1-v^2}r^4}+O(1/r^5)$. So we conclude $\frac{\bar{s}^{r}}{\bar{N}} \bar{\gamma}_{\varphi \varphi} \frac{\partial \bar{\beta}^{\varphi}}{\partial r}\sqrt{\bar{q}} \sim \frac{6am}{\sqrt{1-v^2}} \sin^3\theta$. Then, we also conclude  $\frac{\tilde{s}^{r}}{\tilde{N}} \tilde{\gamma}_{\varphi \varphi}\frac{\partial \tilde{\beta}^{\varphi}}{\partial r} \sqrt{\tilde{q}} \sim \frac{6am}{\sqrt{1-v^2}} \sin^3\theta$ because of $\frac{1}{\Omega^2} \sim 1+O(1/r^2)$. Thus, the Komar angular momentum for (\ref{rometric}) is
\begin{equation}
\tilde{J}=\bar{J}=\frac{1}{16\pi}\int \frac{6 a m}{\sqrt{1-v^{2}}} \sin ^{3} \theta d \theta d \varphi=\frac{m a}{\sqrt{1-v^{2}}};.
\end{equation}
In all, same as the case of static black holes, all the physical quantities of the rotating black holes in the first law of thermodynamics  $dM=TdS+\Omega_{H}dJ+\Phi dQ $ are not changed after conformal transformation. In this sense, they are twin solutions, but they are different in essence.

\section{CONFORMAL INVARIANCE OF PHYSICAL QUANTITIES BEYOND DILATON THEORY}
 So far, we have found that physical quantities in the  first law of thermodynamics remain unchanged  under the conformal transformation for dilaton black hole solutions. By inspecting the procedures for calculating quantities above, one can see that, according to the definition, quantities such as temperature $T$, entropy $S$, angular velocity of black hole $\Omega_{H}$, electric potential $\Phi$ and electric charge $Q$ don't change under the conformal transformations $\Omega^{-1}=\cosh \frac{\phi}{\sqrt{3}}$ regardless of choices of $K_{1}(\phi)$ or $L_{1}(\phi)$. The remaining quantities are mass $M$ and angular momentum $J$, which raises the question of whether they are also invariant under general conformal transformations $\Omega^{-1}=\cosh \frac{\phi}{\sqrt{3}}$ corresponding to theories beyond   dilaton theory?

Indeed, one may deduce from the definition of angular momentum  (\ref{Jdef1}), that the term  $ \frac{s^{r}}{N} \gamma_{\varphi \varphi} \frac{\partial \beta^{\varphi}}{\partial r} \sqrt{q}$ is multiplied by $\frac{1}{\Omega^2}$ under conformal transformations. Therefore, to ensure the convergence of the integration,  $\frac{\bar{s}^{r}}{\bar{N}} \bar{\gamma}_{\varphi \varphi} \frac{\partial \bar{\beta}^{\varphi}}{\partial r}\sqrt{\bar{q}} $ should converge to a constant $a_1$ multiplied by $\sin^3\theta$, i.e. $\frac{\bar{s}^{r}}{\bar{N}} \bar{\gamma}_{\varphi \varphi} \frac{\partial \bar{\beta}^{\varphi}}{\partial r}\sqrt{\bar{q}} \sim a_{1} \sin^3\theta+O(1/r)$. Then the associated angular momentum is
\begin{equation}
\bar{J}=\frac{1}{16\pi}\int a_{1} \sin ^{3} \theta d \theta d \varphi=\frac{a_{1}}{6}\;.
\end{equation}

If the scalar field $\phi$ vanishes at infinity, we obtain  $\Omega^{-1}=cosh(0)=1$ at space infinity. Therefore, according to (\ref{intj1}), we have $\frac{\tilde{s}^{r}}{\tilde{N}} \tilde{\gamma}_{\varphi \varphi}\frac{\partial \tilde{\beta}^{\varphi}}{\partial r} \sqrt{\tilde{q}} \sim  a_{1} sin(\theta)^3+O(1/r) $, indicating that the black hole's angular momentum is conformally invariant  $\tilde{J}=\bar{J}$. By returning to the definition of Komar mass (\ref{mass1}) and concentrating first on the second term of the integrand (\ref{mass2}),   it is trivial to demonstrate that under the  conformal transformation
\begin{equation}
\label{mass3}
\tilde{K}_{r \varphi}\tilde{s}^r\tilde{\beta}^{\varphi}\sqrt{\tilde{q}}=\frac{1}{\Omega^2}\bar{K}_{r \varphi}\bar{s}^r\bar{\beta}^{\varphi}\sqrt{\bar{q}}\;,
\end{equation}
the first term of the integrand in (\ref{mass1}) yields
\begin{equation}
s^iD_iN\sqrt{q}=\sqrt{\frac{g_{22}g_{33}}{g_{11}}}\partial_{r}\sqrt{-\frac{1}{g^{00}}}\;.
\end{equation}
It is not difficult to demonstrate that under conformal transformations, we have
\begin{equation}
\label{mass4}
\tilde{s}^iD_i\tilde{N}\sqrt{\tilde{q}}=\frac{1}{\Omega^2}\bar{s}^iD_i\bar{N}\sqrt{\bar{q}}+\partial_{r}(\frac{1}{2\Omega^2})\bar{s}^r\bar{N}\sqrt{\bar{q}}\;.
\end{equation}

If it is shown that the second term $\partial_{r}(\frac{1}{2\Omega^2})\bar{s}^r\bar{N}\sqrt{\bar{q}}$ makes no contributions to the integral used to define Komar mass,
then combing (\ref{mass3}) and (\ref{mass4}), one obtains
\begin{equation}
\left(\tilde{s}^{i} D_{i} \tilde{N}-\tilde{K}_{i j} s^{i} \tilde{\beta}^{j}\right) \sqrt{\tilde{q}} =\frac{1}{\Omega^2}\left(\bar{s}^{i} D_{i} \bar{N}-\bar{K}_{i j} \bar{s}^{i} \beta^{j}\right) \sqrt{\bar{q}} \;.
\end{equation}
Since we have $\frac{1}{\Omega^2}\sim 1$ at space infinity, the Komar mass (see equation (\ref{mass1}))stays unaltered  $\bar{M}=\tilde{M}$ for the same reason as for conformal invariance of angular momentum. The last task is to demonstrate that $\partial_{r}(\frac{1}{2\Omega^2})\bar{s}^r\bar{N}\sqrt{\bar{q}}$ genuinely doesn't make contributions in the integral. Assuming that the scalar field $\phi$ vanishes and can be expanded into the series in the following form
\begin{equation}
\phi=c_{1}/r+c_{2}/r^2+...\;,
\end{equation}
we find that the conformal transformations can  be expanded into
\begin{equation}
\frac{1}{\Omega^2}=\cosh^2(\frac{\phi}{\sqrt{3}})=1+\frac{c_{1}^2}{3r^2}+O(1/r^3)\;,
\end{equation}
resulting in $\partial_{r}(\frac{1}{2\Omega^2})\sim -\frac{c_{1}^2}{3r^3}+O(1/r^4)$. If  the solution of the black hole for $\mathcal{S}_{1}$ is also asymptotically flat, we obtain  $\bar{s}^r=\sqrt{1/\bar{g}_{11}}\sim 1 ,\bar{N}=\sqrt{-1/\bar{g}^{00}} \sim 1$, $\sqrt{\bar{q}}\sim r^2$.
Thus, when all of the criteria above are satisfied, we get  $\partial_{r}(\frac{1}{2\Omega^2})\bar{s}^r\bar{N}\sqrt{\bar{q}} \sim O(1/r)$, which doesn't contribute to the integral used to define Komar mass. As a result,  the  Komar mass is conformally invariant. Additionally, physical quantities in the first law of thermodynamics $dM=TdS+\Phi dQ +\Omega_{H} dJ$ are conformally invariant.

Why is it necessary that $\phi$ should be expanded into the Taylor series at space infinity? The following counterexample can illustrate this point.
If $\phi$ equals $1/\sqrt{r}$, it cannot be expanded to Taylor series. In this instance, the conformal transformation may  be expanded into
\begin{equation}
\frac{1}{\Omega^2}=1+\frac{1}{3r}+O(1/r^2).
\end{equation}
Under this circumstance, we obtain $\partial_{r}(\frac{1}{2\Omega^2})\bar{s}^r\bar{N}\sqrt{\bar{q}} \sim -\frac{1}{6}$, which does contribution to the integral in definition of Komar mass. Then, we do not have the conclusion of $\bar{M}=\tilde{M}$ in this scenario.

To summarize, the following criteria must be fulfilled in order for all physical quantities in the first law of thermodynamics   to remain conformally invariant regardless of the choices of  $K_1(\phi)$ or $L_1(\phi)$:

1. The spacetime is asymptotically flat so that quantities such as Komar mass and angular momentum can be well defined;

2. At spatial infinity, the scalar field should vanish and be expanded into the Taylor series: $\phi=c_{1}/r+c_{2}/r^2+...\;.$

One can verify that both the static dilaton black hole solution and the precise rotating dilaton black hole solution satisfy the two criteria above, ensuring that the prior analysis of these two solutions is self-consistent.

 \section{Conclusion and Discussion}

In 1974, Bekenstein attempted to obtain exact solutions of conformally invariant scalar equations in Einstein's frame. He established the theorem by which the Einstein-conformal-scalar solution can be deduced from the initial ordinary solution. We take it a step further by taking into account the non-minimal coupling between  the scalar and Maxwell fields, which has generated considerable attention as a consequence of studies on the scalarization of black holes.

The paper starts from the extension of Bekenstein's theorem to EMS theory. Then, using dilaton theory as an illustration of the extended theorem, the conformal dilaton solution is found straightforwardly. Concerning the exact spinning dilaton black hole, its conformal counterpart is obtained and studied. We examine the thermodynamics of our two solutions. We find that the Bekenstein-Hawking entropy formula no longer holds in conformal theory and should be replaced with the Wald entropy formulas,  because the geometry quantity is $(1-\frac{\psi^2}{3})\tilde{R}$ rather than the Ricci scalar $\tilde{R}$. A thorough study shows that the solutions in dilaton theory and their conformal counterparts have the same thermodynamic characteristics and therefore they are cousin models.

After examining these two specific theories, we find that if one can establish that the mass $M$ and angular momentum $J$ are conformally invariant for any conformal factor $\Omega^{-1}=\cosh \frac{\phi}{\sqrt{3}}$ regardless of choices of $K_1(\phi)$ or $L_1(\phi)$, then all the quantities in the first law of thermodynamics $dM=TdS+\Omega_{H}dJ+\Phi dQ$ are conformally invariant. It is therefore found that if the following conditions are satisfied:  (i) the spacetime is asymptotically flat; (ii)
the solution for scalar field in $\mathcal{S}_1$ vanishes at spatial infinity and can be extended into the Taylor series: $\phi=c_{1}/r+c_{2}/r^2+...$, then the physical quantities in the first law of thermodynamics are always conformally invariant.

In terms of potential applications of the extended theorem, one might start with spontaneous scalarization of black holes. One has found that the following three instances \cite{herdeiro2018spontaneous,fernandes2019spontaneous,astefanesei2019einstein}: (i). exponential coupling,
$K_{1}(\phi)=e^{-\alpha \phi^{2}}\;,\ L_{1}(\phi)=0\;;$  (ii). power-law coupling,
$K_{1}(\phi)=1-\alpha \phi^{2}\;,\ L_{1}(\phi))=0\;;$  (iii). fractional coupling,
$K_{1}(\phi)=\frac{1}{1+\alpha \phi^{2}}\;,\ L_{1}(\phi)=0\;;$  allow for spontaneous scalarisation of charged black holes. So it is interesting to investigate the scalarisation of their corresponding conformal version of charged black holes.
Another possible application of the extended theorem is to study black hole perturbations in EMCS theories. The theorem demonstrates that the solutions of $\mathcal{S}_1$ and $\mathcal{S}_2$ are connected via conformal translation regardless of their static, stationary, or even non-stationary properties. Thus the perturbation in EMCS theories can be converted into the investigation on the perturbation in EMS theories.
  Zou et. al \cite{zou2020radial,zou2020scalar} examined the stability problem of scalar hairy black holes in the EMCS theory with the quadratic coupling
 $K_{2}(\psi)=1+\alpha \psi^2\;,\ L_2(\psi)=0\;.$
Indeed, one may perform the stability analysis by using the conformal transformation on the initial solutions in EMS theory. Additionally, Bl{\'a}zquez-Salcedo et.al \cite{blazquez2020einstein} find that, the EMS theory with the higher power-law coupling $K_1(\phi)=1-\alpha \phi^4\;,\ L_{1}(\phi)=0$ exhibits an intriguing two-branch space of scalar hair solutions that coexists with the conventional Reissner-Nordstrom black hole. We consider this character may then be readily extended to its conformal counterpart.  Finally, one can also explore the Einstein-Maxwell theory with an axion coupling \cite{lee1991charged} $K_{1}(\phi)=1\;,\ L_{1}(\phi)=\alpha \phi\;$ by applying the extended theorem.

\section*{ACKNOWLEDGMENTS}
The author Jianhui Qiu would like to express his deepest appreciation to his friend Le Xin Ling, who encourages him a lot and acts like a Patronus in \emph{Harry potter}. Additionally, he wishes to express his gratitude to Prof. Gao for his patience and assistance.
This work is partially supported by the NSFC under grants 11633004 and 11773031.

\section*{REFERENCES}

\bibliography{citationlist.bib}

\end{document}